\documentclass{article}%
\usepackage{amsmath}
\usepackage{graphicx}
\usepackage{amsfonts}
\usepackage{amssymb}%
\providecommand{\U}[1]{\protect\rule{.1in}{.1in}}
\providecommand{\U}[1]{\protect\rule{.1in}{.1in}}
\begin{document}

\title{ Gravitation and Electromagnetism as Geometrical Objects of a Riemann-Cartan
Spacetime Structure}
\author{J. Fernando T. Giglio$^{(1)}$ and Waldyr A. Rodrigues Jr.$^{(2)}$.\\$^{(1)}$ {\footnotesize FEA-CEUNSP. 13320-902 - Salto, SP, Brazil.}\\{\footnotesize jfernandotg@gmail.com}\\$^{(2)}\hspace{-0.05cm}${\footnotesize Institute of Mathematics, Statistics
and Scientific Computation}\\{\footnotesize IMECC-UNICAMP}\\{\footnotesize 13083-859 Campinas, SP, Brazil}\\{\footnotesize walrod@ime.unicamp.br or walrod@mpc.com.br}}
\maketitle

\begin{abstract}
In this paper we first show that any coupled system consisting of a
gravitational plus a \textit{free} electromagnetic field can be described
\textit{geometrically} in the sense that both Maxwell equations and Einstein
equation having as source term the energy-momentum of the electromagnetic
field can be derived from a geometrical Lagrangian proportional to the scalar
curvature $R$ of a \textit{particular} kind of Riemann-Cartan spacetime
structure. In our model the gravitational and electromagnetic fields are
identified as \emph{geometrical objects} of the structure. We show moreover
that the contorsion tensor of the\ particular Riemann-Cartan spacetime
structure of our theory encodes the same information as the one contained in
\emph{Chern-Simons }term $A\wedge dA$ that is proportional to the spin density
of the electromagnetic field. Next we show that by adding to the geometrical
Lagrangian a term describing the interaction of a electromagnetic current with
a general electromagnetic field plus the gravitational field, together with a
term describing the matter carrier of the current we get Maxwell equations
with source term and Einstein equation having as source term the sum of the
energy-momentum tensors of the electromagnetic and matter terms. Finally
modeling by \textit{dust charged matter} the carrier of the electromagnetic
current we get the Lorentz force equation. Moreover, we prove that our theory
is \textit{gauge invariant}. We also briefly discuss our reasons for the
present enterprise.

\end{abstract}

\section{Introduction}

Since the geometrization of gravitation by General Relativity (GR) where\ the
gravitational field, generated by an energy-momentum tensor\footnote{In this
paper the notation $\sec T_{s}^{r}M$ means section the the $T_{s}^{r}M$
bundle. Also, $\sec%
{\textstyle\bigwedge\nolimits^{r}}
T^{\ast}M$ means section of the bundle of $r$-form fields.} $\boldsymbol{T\in
}\sec T_{0}^{2}M$, is represented by a particular Lorentzian spacetime
structure\footnote{In the structure $\langle M,\boldsymbol{g},\bar{\nabla
},\tau_{\boldsymbol{g}},\uparrow\rangle$, the pair $\langle M,\boldsymbol{g}%
\rangle$ is called a Lorentzian manifold, $M$ being a $4$-dimensional
Hausdorff paracompact locally compact manifold and $\boldsymbol{g\in}\sec
T_{0}^{2}M$ a Lorentzian metric of signature $(1,-1-,1-1)$. $\bar{\nabla}$ is
the Levi-Civita connection of $\boldsymbol{g}$, $\tau_{\boldsymbol{g}}\in\sec%
{\textstyle\bigwedge\nolimits^{4}}
T^{\ast}M$ and $(\uparrow)$ define respectively a spacetime orientation and a
time orientation for $M$. More details about time orientation may be found,
e.g., in \cite{sawu}.} $\langle M,\boldsymbol{g},\bar{\nabla},\tau
_{\boldsymbol{g}},\uparrow\rangle$ together with Einstein equation, several
classical models have been proposed which try to geometrize the description of
the electromagnetic field, with the obvious interest in describing both
fields, i.e., gravitational and electromagnetic, through a unique principle:
the geometric one.

This has been tried by generalizing the Lorentzian spacetime structure, i.e.,
utilizing more general geometries incorporating additional degrees of freedom,
which hopefully permits in principle a description of the electromagnetic
field as some aspect of the underlie chosen geometry. In this sense, there has
been many lines of investigation of this problem, some of the most well known are:

(i) Weyl theory, where non metric compatible symmetric connections \cite{1,2}
are used. The resulting geometry has non zero Riemann tensor and a null
torsion tensor and is now known as Weyl geometries \cite{3}. We mention also
in this class Eddington unified theory which is described by a non metric
compatible connection with non null Riemann and torsion tensors \cite{4}.

(ii) Theories based on spacetimes with more than four dimensions \cite{5,6},
known as Kaluza-Klein theories. These theories have been studied in the last
decades in connection with fiber bundle formulations of the four fundamental
interactions \cite{7,8}.

(iii) Introduction of metric compatible connections in four dimensions, other
than that the Levi-Civita connection, like, e.g., in the \textbf{so-called}
Riemann-Cartan geometries \cite{9,23}, which in general have non zero Riemann
and torsion tensors.

(iv) The non symmetric metric theory of Einstein \cite{17}.

(v) Theories on Finslerian spaces \cite{18,19,20,21}.

(vi) Einstein teleparallel theory\footnote{See complete list of Einstein
papers on teleparallelism in \cite{sauer}.}.

The fact is that all these theories are problematic. According to the majority
view, Weyl theory received what is thought to be a knockdown by Einstein
\cite{22} but according to Eddington, non metricity can not be completely
ruled out by experiment if due care is taken \cite{4}. Concerning (ii) it
remains always a problem to explain why the extra dimensions are not
observable, or why they compactify\footnote{The problem exists also in modern
string theory, for no reason is given for the compatification of the extra
dimensions.} \cite{8}. Concerning (iii), several theories that includes
Cartan's torsion in GR have been proposed. One motivation was to obtain a
unified geometrical description of electromagnetism and gravitation
\cite{18,25}. These theories, and that of (i) and (ii), at least to what
refers to a classical unified description of the gravitational and
electromagnetic fields, are not in general totally accepted because of their
failure in obtaining simultaneously the electromagnetic field equations
(Maxwell equations), the energy-momentum tensor of the electromagnetic field
and the Lorentz force equation for the motion of charged matter as they are
known in the physical situations involving gravitation and electromagnetism in
a four dimensional universe. Also, some authors\footnote{Well, this is indeed
a polemical view, not endorsed, e.g., by Weinberg. See his exchange of letters
with Hehl in \emph{Physics Today} at:
http://ptonline.aip.org/journals/doc/PHTOAD-ft/vol\_60/iss\_3/16\_2.shtml?bypassSSO=1.
Anyway there are some proposals in the literature to observe experimentally
the existence of torsion, see, e.g., \cite{37,35}.} are of the opinion that
Riemann-Cartan geometry is necessary to describe besides gravitation, also the
spinning matter, which is supposed to be the source of \ the torsion field
\cite{23,24,25,25aa,25a} and it seems also that a non vanishing torsion tensor
appears as a necessary ingredient in the gauge formulation of GR when the
Poincar\'{e} invariance is taken locally \cite{25,26,27,28,29}. Besides that
in string theory, in the low energy limit, the effective Lagrangian has an
antisymmetric field that is interpreted as torsion \cite{30}. Torsion derived
from scalar field, vector field or antisymmetric tensor field can be found in
Hammond \cite{31} and references cited therein. The non-symmetric theory of
Einstein has recently\ be developed by Moffat \cite{12} with a very different
interpretation aiming to describe dark matter, but will not be commented here,
nor will we discuss the status of theories that use Finslerian spacetimes.
However we comment that Einstein's teleparallel theory, that he originally
interpreted as a unified theory of the gravitational plus the electromagnetic
field\ is indeed a non sequitur. The case is that Einstein's preferred version
of his teleparallel theory has a Lagrangian that is equivalent to the
Einstein-Hilbert Lagrangian of GR\footnote{This is clear, e.g., in \ \cite{32}
where the torsion tensor is used to describe the gravitational field of
Einstein's GR in a\ particular Riemann-Cartan spacetime structure known
nowadays as Weitzenb\"{o}ck (or teleparallel) spacetime $\langle
M,\boldsymbol{g},\nabla,\tau_{\boldsymbol{g}},\uparrow\rangle$ where the
curvature tensor of $\nabla$ is null, and the torsion tensor of $\nabla$ is
non null. See also \cite{rod2011}.}. This according to \cite{sauer} has been
informed by Lancoz to Einstein and put, so to say, an end to teleparallelism
as an unified field theory

Moreover, it is now known \cite{rod2011} that a theory of the gravitational
field can be formulated for the gravitational potentials $\mathfrak{g}%
^{\mathbf{a}}\in\sec%
{\textstyle\bigwedge\nolimits^{1}}
T^{\ast}M$ (with at least one of the $\mathfrak{g}^{\mathbf{a}}$ for
$a=0,1,2,3$ non closed, i.e., $\boldsymbol{F}^{\mathbf{a}}=d\mathfrak{g}%
^{\mathbf{a}}\neq0$) living on Minkowski spacetime $\langle M,\boldsymbol{\eta
},D,\tau_{\mathbf{\eta}},\uparrow\rangle$ and satisfying field equations
derived from a postulated Lagrangian density, thus dispensing the geometrical
interpretation of gravitation as a Lorentzian or a teleparallel spacetime. The
field equations of the theory are easily seem to be equivalent to Einstein's
equations once we introduce a field $\boldsymbol{g}=\eta_{\mathbf{ab}%
}\mathfrak{g}^{\mathbf{a}}\otimes\mathfrak{g}^{\mathbf{b}}\in\sec T_{0}^{2}M$
together with its Levi-Civita connection $\bar{\nabla}$ and interpret the
structure $\langle M,\boldsymbol{g},\bar{\nabla},\tau_{\boldsymbol{g}%
},\uparrow\rangle$ as an \emph{effective} Lorentzian spacetime\footnote{There
are other possibilities also involving more complicated geometrical
structures, see, e.g., \cite{nrr2010}.}.

It is also the case that for the formulation of the electromagnetic field
theory we do not even need a metric field defined on a manifold that serves as
support for that field \cite{helobu}. Indeed, it is well know that Maxwell
equations can be written in a \textit{star shape} manifold \cite{choquet} as
the compatibility equations for a closed $2$-form field $\boldsymbol{F}\in\sec%
{\textstyle\bigwedge\nolimits^{2}}
T^{\ast}M$ ($d\boldsymbol{F}=0$) and a closed current $\mathbf{J}\in\sec%
{\textstyle\bigwedge\nolimits^{3}}
T^{\ast}M$ ($d\mathbf{J}=0$). These equations imply in a star shape manifold
the existence of a $1$-form field $\boldsymbol{A}\in\sec%
{\textstyle\bigwedge\nolimits^{1}}
T^{\ast}M$ and a $2$-form field $\boldsymbol{G}\in\sec%
{\textstyle\bigwedge\nolimits^{2}}
T^{\ast}M$ such that $\boldsymbol{F}=d\boldsymbol{A}$ and $\mathbf{J=-}%
d\boldsymbol{G}$ in such a way that Maxwell equations read
\begin{equation}
d\boldsymbol{F}=0,\text{ \ }d\boldsymbol{G}=\mathbf{-J} \label{me}%
\end{equation}

It is also well known that in general $\boldsymbol{G}$ is related \ to
$\boldsymbol{F}$ through the \textbf{so-called} constitutive equations of the
medium. In that sense the gravitational field in GR modeled by a Lorentzian
spacetime serves as an effective medium for the propagation of the
electromagnetic field \ and the constitutive equations are given simple by
\begin{equation}
\boldsymbol{G}=\underset{g}{\star}\boldsymbol{F,}\label{me1}%
\end{equation}
\ and thus defining $\boldsymbol{J=}\underset{g}{\star}\mathbf{J\in}\sec%
{\textstyle\bigwedge\nolimits^{1}}
T^{\ast}M$ we can write Maxwell equations as
\begin{equation}
d\boldsymbol{F}=0,\text{\ \ }\delta\boldsymbol{F}=-\boldsymbol{J},\label{me2}%
\end{equation}
where $\underset{g}{\star}$ is the Hodge dual operator and $\delta$ is the
Hodge coderivative operator\footnote{Details, may be found, e.g., in
\cite{rodcap2007}.}. The intrinsic Maxwell equations even when expressed in a
Lorentzian or Riemann-Cartan spacetime structures do not need the use of the
covariant derivative operator of those structures for their writing, although
they can be formulated with such operators (see below\footnote{See also
\cite{rodflb} for the formulation of Maxwell equations using the the covariant
derivative operator of a general Riemann-Cartan spacetime structure.}$^{,}%
$\footnote{Moreover it can be shown that at least in Minkowski spacetime those
equations and energy-momentum conservation of field plus matter imply in a
unique coupling between $\boldsymbol{F}$ and $\boldsymbol{J}$, namely the
Lorentz force law \cite{rr2010}.}).

Having said all that we can ask: is it possible, in the same sense that the
gravitational field can be describe by the potentials $\mathfrak{g}%
^{\mathbf{a}}$ living in Minkowski spacetime or by a Lorentzian structure
$\langle M,\boldsymbol{g},\bar{\nabla},\tau_{\boldsymbol{g}},\uparrow\rangle$
or a teleparallel structure $\langle M,\boldsymbol{g},\nabla,\tau
_{\boldsymbol{g}},\uparrow\rangle$ to describe gravitation and
electromagnetism geometrically, i.e., through a particular Riemann-Cartan
spacetime structure where those fields are represented by some of the
geometrical objects associated with that structure?

As we shall see, the answer is positive once we use a Riemann-Cartan structure
equipped with a particular connection whose contortion tensor is given by
Eq.(\ref{20}) below\footnote{We observe that in many of the papers dealing
with the subject of our study torsion is generally taken in particular forms.
Besides that let us also add that torsion as resulting from spinning matter
has never been experimentally observed \cite{33,34}, although there are
proposals to this end \cite{35,37}.}.

In our very simple model we are able to obtain a unified description of the
gravitational and \textit{free }electromagnetic fields as geometrical aspects
of a particular Riemann-Cartan spacetime, with the Einstein and the free
Maxwell equations being derived from a geometrical Lagrangian, i.e., a
Lagrangian proportional to the scalar curvature $R$ of a particular
Riemann-Cartan connection. The main mathematical tools for doing that is
presented in Section 2 where we also show that the information contained in
the contortion tensor of the particular Riemann-Cartan spacetime structure of
our theory is the same as the one contained in the Chern-Simons term $A\wedge
dA$ that as well known now \cite{rodcap2007,devries} is proportional to the
spin density of the electromagnetic field.

Moreover, we show in Section 3 that by adding to the geometrical Lagrangian an
interaction term proportional to $\boldsymbol{J}\cdot\boldsymbol{A}$
describing the source of the electromagnetic\ field and its interaction with
that field and a term describing the matter carrier of the current we get
Maxwell equations with source term and Einstein equations having as source
term the sum of the energy-momentum tensors of the electromagnetic and matter
terms. In Section 4, modeling by \textit{dust charged matter} the carrier of
the electromagnetic current we get from the nullity of Riemann-Cartan
covariant derivative of the sum of energy-momentum tensor of matter plus the
electromagnetic field the Lorentz force equation. In Section 5 we show that
our theory is well defined by proving its gauge invariance. Finally, in
Section 6 we present our conclusions.

\section{Maxwell Equations}

In what follows $\langle M,\boldsymbol{g}\rangle$ as defined above is a
Lorentzian manifold. Let $\bar{\nabla}$ and $\nabla$ be respectively the
Levi-Civita connection and a particular metric compatible Riemann-Cartan
\cite{36} connection of $\boldsymbol{g}$ on $M$. Let $U\subset M$ and $\langle
x^{\mu}\rangle$ be a coordinates for $U\subset$ $M$, and $\langle
\boldsymbol{e}_{\mu}=\partial/\partial x^{\mu}\rangle$ a basis of $TU$
$(\mu=0,1,2,3)$ and $\langle\vartheta^{\mu}=dx^{\mu}\rangle$ the corresponding
dual basis \ i.e., a basis for $T^{\ast}U$. \ We also introduce the reciprocal
basis $\langle\boldsymbol{e}^{\mu}\rangle$\ of $\langle\boldsymbol{e}_{\mu
}\rangle$ for $TM$ and the reciprocal basis $\langle\vartheta_{\mu}\rangle$ of
$\langle\vartheta^{\mu}\rangle$ for $T^{\ast}M$, such that%
\begin{align}
\boldsymbol{g} &  =g_{\mu\nu}\vartheta^{\mu}\otimes\vartheta^{\nu}=g^{\mu\nu
}\vartheta_{\mu}\otimes\vartheta_{\nu},\text{ \ \ }g^{\mu\alpha}g_{\alpha\nu
}=\delta_{\nu}^{\mu},\nonumber\\
\boldsymbol{e}^{\mu} &  =g^{\mu\nu}\boldsymbol{e}_{\nu},\text{ \ \ }%
\vartheta_{\mu}=g_{\mu\nu}\vartheta^{\nu}.\label{rec}%
\end{align}
Moreover we introduce as metric for the cotangent bundle the object $g\in\sec
T_{2}^{0}M,$%
\[
g=g^{\mu\nu}\boldsymbol{e}_{\mu}\otimes\boldsymbol{e}_{\nu}=g_{\mu\nu
}\boldsymbol{e}^{\mu}\otimes\boldsymbol{e}^{\nu}%
\]
\ and define the scalar product of arbitrary $1$-form fields $\boldsymbol{X}$
and $\boldsymbol{Y}$ by
\begin{equation}
\boldsymbol{X}\cdot\boldsymbol{Y}=g(\boldsymbol{X},\boldsymbol{Y})\label{SP}%
\end{equation}
Let moreover $\Gamma_{\mu\nu\cdot}^{\cdot\cdot\lambda}$ and $\bar{\Gamma}%
_{\mu\nu\cdot}^{\cdot\cdot\lambda}$ be the connection coefficients of $\nabla$
and $\bar{\nabla}$ in the coordinate basis just introduced, i.e.,
$\nabla_{\partial_{\mu}}\partial_{\nu}=\Gamma_{\mu\nu\cdot}^{\cdot\cdot
\lambda}\partial_{\lambda}$ and $\bar{\nabla}_{\partial_{\mu}}\partial_{\nu
}=\bar{\Gamma}_{\mu\nu\cdot}^{\cdot\cdot\lambda}\partial_{\lambda}$. As it is
well know \ (see, e.g., \cite{38,rodcap2007}), $\bar{\Gamma}_{\mu\nu\cdot
}^{\cdot\cdot\lambda}$ and $\Gamma_{\mu\nu\cdot}^{\cdot\cdot\lambda}$ are
related by
\begin{equation}
\Gamma_{\mu\nu\cdot}^{\cdot\cdot\lambda}=\bar{\Gamma}_{\mu\nu\cdot}%
^{\cdot\cdot\lambda}+K_{\mu\nu\cdot}^{\cdot\cdot\lambda\hspace{0.01in}%
}\text{,}\label{1}%
\end{equation}
where the connection coefficients $\Gamma_{\mu\nu\cdot}^{\cdot\cdot\lambda}$
of the Levi-Civita connection are given by:
\begin{equation}
\Gamma_{\mu\nu\cdot}^{\cdot\cdot\lambda}=\frac{1}{2}g^{\lambda\alpha}\left(
\partial_{\mu}g_{\nu\alpha}+\partial_{\nu}g_{\mu\alpha}-\partial_{\alpha
}g_{\mu\nu}\right)  \label{2}%
\end{equation}
and the $K_{\mu\nu}^{\cdot\cdot\lambda\hspace{0.01in}}$ are the components of
the contorsion tensor $\mathcal{K}=K_{\mu\nu\cdot}^{\cdot\cdot\beta}e_{\beta
}\otimes dx^{\mu}\otimes dx^{\nu}\in TM\otimes\sec T_{1}^{2}M$ defined by
\footnote{Note that this differs from the definition in \cite{24} by a signal
and a factor $1/2$ due to conventions.used here with are the ones in
\cite{rodcap2007}.}:%

\begin{align}
K_{\mu\nu\cdot}^{\cdot\cdot\beta}  &  :=\frac{1}{2}(g^{\lambda\beta}%
g_{\lambda\rho}T_{\mu\nu\cdot}^{\cdot\cdot\rho}-g^{\lambda\beta}g_{\nu\rho
}T_{\mu\lambda\cdot}^{\cdot\cdot\rho}-g^{\lambda\beta}g_{\mu\rho}T_{\nu
\lambda\cdot}^{\cdot\cdot\rho})\nonumber\\
&  =\frac{1}{2}(T_{\mu\nu\cdot}^{\cdot\cdot\hspace{0.01in}\beta}-T_{\mu
\cdot\nu}^{\cdot\beta\cdot}+T_{\cdot\nu\mu}^{\beta\cdot\cdot}). \label{cont}%
\end{align}
However, taking into account that we have the (bastard \cite{gs}) symmetry
$K_{\mu\nu\lambda}^{\cdot\cdot\cdot}:=g_{\beta\lambda}K_{\mu\nu\cdot}%
^{\cdot\cdot\beta}=-K_{\mu\lambda\nu}^{\cdot\cdot\cdot}$ we prefer in what
follows to take the contortion as the object $\boldsymbol{K\in}%
{\textstyle\bigwedge\nolimits^{1}}
T^{\ast}M\otimes%
{\textstyle\bigwedge\nolimits^{2}}
T^{\ast}M$ (which carries the same information as $\mathcal{K}$) defined by
\begin{equation}
\boldsymbol{K}=\frac{1}{2}K_{\mu\nu\cdot}^{\cdot\cdot\lambda}\vartheta^{\mu
}\otimes\vartheta^{\nu}\wedge\vartheta_{\lambda}=\frac{1}{2}K_{\mu\nu\lambda
}^{\cdot\cdot\cdot}\vartheta^{\mu}\otimes\vartheta^{\nu}\wedge\vartheta
^{\lambda}. \label{9}%
\end{equation}

Also, the
\begin{equation}
T_{\mu\nu\cdot}^{\cdot\cdot\lambda}=\left(  \Gamma_{\mu\nu\cdot}^{\cdot
\cdot\lambda}-\Gamma_{\nu\mu\cdot}^{\cdot\cdot\lambda}\right)  \label{3}%
\end{equation}
are the components of torsion tensor $\mathit{\Theta}=\frac{1}{2}T_{\mu
\nu\cdot}^{\cdot\cdot\lambda}\boldsymbol{e}_{\lambda}\otimes\vartheta^{\mu
}\wedge\vartheta^{\nu}$ $\in\sec TM\otimes%
{\textstyle\bigwedge\nolimits^{2}}
T^{\ast}M$ of the Riemann-Cartan connection $\nabla$, but here we will prefer
to use as torsion tensor the object \ $\boldsymbol{\Theta}=\frac{1}{2}%
T_{\mu\nu\cdot}^{\cdot\cdot\lambda}\vartheta_{\lambda}\otimes\vartheta^{\mu
}\wedge\vartheta^{\nu}$ $\in\sec%
{\textstyle\bigwedge\nolimits^{1}}
TM\otimes%
{\textstyle\bigwedge\nolimits^{2}}
T^{\ast}M$ which encodes the same information than $\Theta.\medskip$

We now proceed by introducing a \textit{particular }Riemann-Cartan spacetime
structure $\langle M,\boldsymbol{g},\nabla,\tau_{\boldsymbol{g}}%
,\uparrow\rangle$ where the contorsion tensor is defined by%
\begin{equation}
\boldsymbol{K}:=-C\boldsymbol{B\otimes}\boldsymbol{F}\in\sec%
{\textstyle\bigwedge\nolimits^{1}}
T^{\ast}M\otimes%
{\textstyle\bigwedge\nolimits^{2}}
T^{\ast}M
\end{equation}
with components\footnote{This form is analogous to such that is taken in
\cite{39} for torsion.}%
\begin{equation}
K_{\mu\nu\cdot}^{\cdot\cdot\lambda}=-CB_{\mu}F_{\nu\cdot}^{\cdot\lambda
},\text{ \ \ }K_{\mu\nu\lambda}^{\cdot\cdot\cdot}=-CB_{\mu}F_{\nu\lambda
}=-K_{\mu\lambda\nu}^{\cdot\cdot\cdot}. \label{K}%
\end{equation}
and where the constant $C$ and$\ $the $B_{\mu}$, which are the components of
$\boldsymbol{B\in\sec}%
{\textstyle\bigwedge\nolimits^{1}}
T^{\ast}M$, are to be determined and where the $F_{\nu\cdot}^{\cdot\lambda
}:=g^{\lambda\alpha}F_{\nu\alpha}=-g^{\alpha\lambda}F_{\alpha\nu}=-F_{\cdot
\nu}^{\lambda\cdot}$ are the components of $\boldsymbol{F}\in\sec%
{\textstyle\bigwedge\nolimits^{2}}
T^{\ast}M$, i.e.,
\begin{equation}
\boldsymbol{F}:=\frac{1}{2}F_{\mu\nu}\vartheta^{\mu}\wedge\vartheta^{\nu
}=\frac{1}{2}F_{\mu\cdot}^{\cdot\nu}\vartheta^{\mu}\wedge\vartheta_{\nu}.
\label{4a}%
\end{equation}
In what follows we propose to give a physical interpretation for those objects
associated to the structure $\langle M,\boldsymbol{g},\nabla,\tau
_{\boldsymbol{g}},\uparrow\rangle$.

Before proceeding to build our theory we recall some formulas that will be
used latter. We start calculating the covariant derivative $\nabla_{\mu}$ of
$F^{\mu\nu}$ \footnote{More precisely we write , e.g., for a tensor
$t=t_{\alpha}^{\cdot\nu}\partial_{\nu}\otimes dx^{\alpha}\in\sec TM\otimes
T^{\ast}M$, $\nabla_{\partial_{\mu}}t=(\nabla_{\mu}t_{\alpha}^{\cdot\nu
})\partial_{\nu}\otimes dx^{\alpha}$ where $\nabla_{\mu}t_{\alpha}^{\cdot\nu
}=\partial_{\mu}t_{\alpha}^{\cdot\nu}+\Gamma_{\mu\delta}^{\cdot\cdot\nu
}t_{\alpha}^{\cdot\delta}-\Gamma_{\mu\alpha}^{\cdot\cdot\delta}t_{\delta
}^{\cdot\nu}$.}. Using the connection given by Eq.(\ref{1}) we obtain
\begin{equation}
\nabla_{\mu}F^{\mu\nu}=\bar{\nabla}_{\mu}F^{\mu\nu}+K_{\mu\delta}^{\cdot
\cdot\mu}F^{\delta\nu}+K_{\mu\delta}^{\cdot\cdot\nu}F^{\mu\delta}\text{.}
\label{5}%
\end{equation}

But the last two terms in Eq.(\ref{5}) cancel out due to Eq.(\ref{K}), and we
have
\begin{equation}
\nabla_{\mu}F^{\mu\nu}=\bar{\nabla}_{\mu}F^{\mu\nu}=\frac{1}{(-\det
g)^{\frac{1}{2}}}\partial_{\mu}[(-\det g)^{\frac{1}{2}}F^{\mu\nu}]\text{.}
\label{6}%
\end{equation}

Next, we observe that from Eq.(\ref{5}) it follows immediately that
\begin{align}
\nabla_{\lbrack\mu}F_{\nu\lambda]}  &  =\nabla_{\mu}F_{\nu\lambda}+\nabla
_{\nu}F_{\lambda\mu}+\nabla_{\lambda}F_{\mu\nu}=\bar{\nabla}_{[\mu}%
F_{\nu\lambda]}\nonumber\\
&  =\partial_{\mu}F_{\nu\lambda}+\partial_{\nu}F_{\lambda\mu}+\partial
_{\lambda}F_{\mu\nu}. \label{12}%
\end{align}

From Eqs.(\ref{6}) and (\ref{12}), a natural assumption is to define
$\boldsymbol{F}=d\boldsymbol{A}$, where $\boldsymbol{A}\in\sec%
{\textstyle\bigwedge\nolimits^{1}}
T^{\ast}M$, and of course,%
\begin{equation}
F_{\mu\nu}=\partial_{\mu}A_{\nu}-\partial_{\nu}A_{\mu}=\bar{\nabla}_{\mu
}A_{\nu}-\bar{\nabla}_{\nu}A_{\mu}\text{.} \label{13}%
\end{equation}
This suggests to interpret $\boldsymbol{F}$ as the electromagnetic field and
$\boldsymbol{A}$ as its potential, and we are going to show that this is
indeed the case.

Also, Eq.(\ref{6}) defines in general a conserved current $1$-form field
$\boldsymbol{J}=J_{\mu}\vartheta^{\mu}=J^{\nu}\vartheta_{\nu}\in\sec%
{\textstyle\bigwedge\nolimits^{1}}
T^{\ast}M$ such that ($c$ being the velocity of light in vacuum)
\begin{equation}
\frac{4\pi}{c}J^{\nu}:=\nabla_{\mu}F^{\mu\nu}=\bar{\nabla}_{\mu}F^{\mu\nu
}=(-\det\boldsymbol{g})^{-\frac{1}{2}}\partial_{\mu}[(-\det\boldsymbol{g}%
)^{\frac{1}{2}}F^{\mu\nu}] \label{15}%
\end{equation}
and of course,
\begin{equation}
\partial_{\nu}[(-\det\boldsymbol{g})^{\frac{1}{2}}J^{\nu}]=0. \label{16}%
\end{equation}

With our choise of $\boldsymbol{F}$ the second member of Eq.(\ref{12}) is
null. We then recognize Eq.(\ref{12}) and Eq.(\ref{15}) as Maxwell equations
written on a Lorentzian spacetime \cite{40,rodflb}.

\section{Action Principle, Maxwell and Einstein\newline Equations and their
Source Terms}

In a Riemann-Cartan spacetime the curvature tensor can be written as (see,
e.g., \cite{38,rodcap2007})
\begin{equation}
R_{\mu\nu\lambda\cdot}^{\cdot\cdot\cdot\chi}=\bar{R}_{\mu\nu\lambda\cdot
}^{\cdot\cdot\cdot\chi}+\bar{\nabla}_{\mu}K_{\nu\lambda\cdot}^{\cdot\cdot\chi
}-\bar{\nabla}_{\nu}K_{\mu\lambda\cdot}^{\cdot\cdot\chi}+K_{\nu\lambda\cdot
}^{\cdot\cdot\rho}K_{\mu\rho\cdot}^{\cdot\cdot\chi}-K_{\mu\lambda\cdot}%
^{\cdot\cdot\rho}K_{\nu\rho\cdot}^{\cdot\cdot\chi}\text{ ,} \label{17}%
\end{equation}
where the bars as already said above refers to quantities defined with the
Levi-Civita connection. The last two terms in Eq.(\ref{17}) cancel out because
of Eq.(\ref{K}) and then from Eq.(\ref{17}) we have for the scalar curvature:
\begin{equation}
R=g^{\nu\lambda}R_{\mu\nu\lambda\cdot}^{\cdot\cdot\cdot\mu}=\bar{R}%
+2\bar{\nabla}_{\mu}K_{\nu\cdot\cdot}^{\cdot\nu\mu}. \label{18}%
\end{equation}
Doing the evaluation of $\bar{\nabla}_{\mu}K_{\nu\cdot\cdot}^{\cdot\nu\mu}$
and taking into account that $\bar{\nabla}_{\mu}B_{\nu}-\bar{\nabla}_{\nu
}B_{\mu}=\partial_{\mu}B_{\nu}-\partial_{\nu}B_{\mu}$ we get
\begin{equation}
R=\bar{R}+C\left(  \partial_{\mu}B_{\nu}-\partial_{\nu}B_{\mu}\right)
F^{\mu\nu}+\frac{8\pi C}{c}B_{\mu}J^{\mu}\text{.} \label{19}%
\end{equation}

Eq.(\ref{19}) suggests to us to identify the $B_{\mu}$ in $K_{\mu\nu\cdot
}^{\cdot\cdot\lambda}$ with\ the components of the electromagnetic potential
$\boldsymbol{A}=A_{\mu}dx^{\mu}$, i.e., we take from now on
\begin{equation}
K_{\mu\nu\cdot}^{\cdot\cdot\lambda}:=-CA_{\mu}F_{\nu\cdot}^{\cdot\lambda
}\text{,} \label{20}%
\end{equation}
since this identification will permit us to interpret (a factor apart) the
fist and second terms on the r.h.s. of Eq.(\ref{19}) as the
gravitational\ \ field and the \textit{free} electromagnetic field ($J_{\mu
}=0)$ Lagrangian in GR. Indeed, for that case we are in position of
interpreting those fields as parts of a Riemann-Cartan spacetime structure
$\langle M,\boldsymbol{g}$,$\nabla,\tau_{\boldsymbol{g},}\uparrow\rangle$ by
taking as Lagrangian of the system
\begin{equation}
L:=\frac{-c^{3}}{16\pi G}R \label{20X}%
\end{equation}
and the action is%
\begin{equation}
S=\frac{-c^{3}}{16\pi G}\int(\bar{R}+CF_{\mu\nu}F^{\mu\nu})(-\det
\boldsymbol{g})^{\frac{1}{2}}d^{4}x. \label{20Y}%
\end{equation}

We can now give a geometrical model for the interaction\ of the
electromagnetic field $\boldsymbol{F}$, its current $\boldsymbol{J}$ and the
gravitational field $\boldsymbol{g}$ by interpreting those fields as parts of
a Riemann-Cartan spacetime structure $\langle M,\boldsymbol{g}$,$\nabla
,\tau_{\boldsymbol{g},}\uparrow\rangle$. Variation (
$\underset{g}{\boldsymbol{\delta}}$ ) of $S$ in Eq.(\ref{20Y}) with respect to
the contravariant components of $\boldsymbol{g}$ gives%
\begin{align}
\underset{g}{\boldsymbol{\delta}}S  &  =\frac{-c^{3}}{16\pi G}%
\underset{g}{\delta}[\int\bar{R}(-g)^{\frac{1}{2}}d^{4}x+C\int F_{\mu\nu
}F^{\mu\nu}(-\det\boldsymbol{g})^{\frac{1}{2}}d^{4}x\label{20Z}\\
&  =\frac{-c^{3}}{16\pi G}\int(\bar{R}_{\mu\nu}-\frac{1}{2}g_{\mu\nu}\bar
{R}-8\pi CT_{\mu\nu})\delta g^{\mu\nu}(-\det\boldsymbol{g})^{\frac{1}{2}}%
d^{4}x,
\end{align}
where
\begin{equation}
T_{\mu\nu}=\frac{1}{4\pi}(-F_{\mu}^{\ \beta}F_{\nu\beta}+\frac{1}{4}g_{\mu\nu
}F^{\alpha\beta}F_{\alpha\beta})\text{,}%
\end{equation}
are the components of the energy-momentum tensor of electromagnetic
electromagnetic field, showing moreover that we must take $C$ as
\begin{equation}
C=\frac{G}{c^{4}}. \label{27}%
\end{equation}

\noindent\textbf{Remark 1 }\emph{This complete the proof of our claim in the
introduction that any coupled system consisting of a gravitational field and
electromagnetic field can be fully geometrized by a special Riemann-Cartan
spacetime structure.\medskip}

\noindent\textbf{Remark 2 }\emph{We observe that if the contorsion tensor just
introduced and whose information is contained in }$\boldsymbol{K}$\emph{ is
the same as the one contained in the Chern-Simons object }\cite{36,cg}\emph{
}$\boldsymbol{A}\wedge d\boldsymbol{A}\in\sec%
{\textstyle\bigwedge\nolimits^{3}}
T^{\ast}M$\emph{. Indeed,}%

\begin{align}
\boldsymbol{C}  &  =\boldsymbol{A}\wedge d\boldsymbol{A=A}\wedge
\boldsymbol{F}\nonumber\\
&  =\frac{1}{2}A_{\mu}F_{\nu\lambda}\vartheta^{\mu}\wedge\vartheta^{\nu}%
\wedge\vartheta^{\lambda}\nonumber\\
&  =\frac{1}{3!}(A_{\mu}F_{\nu\lambda}+A_{\lambda}F_{\mu\nu}+A_{\nu}%
F_{\lambda\mu})\vartheta^{\mu}\wedge\vartheta^{\nu}\wedge\vartheta^{\lambda}
\label{chern-simons}%
\end{align}

\emph{It is eventually opportune to observe that }$\boldsymbol{A}\wedge
d\boldsymbol{A}$\emph{ has been called in }\cite{kiehn}\emph{ the topological
torsion, although in }\cite{rodcap2007}\emph{\ it has been argued that this
was not a good nomenclature since this object is proportional to the spin
density of the electromagnetic field. This results seems\ to be endorsed by
the nice analysis in }\cite{devries}.

\noindent\textbf{Remark 3 }\emph{Before proceeding we note that \cite{42}
}$\underset{g}{\boldsymbol{\delta}}\int A_{\mu}J^{\mu}(-\det\boldsymbol{g}%
)^{\frac{1}{2}}d^{4}x=0$\emph{. Indeed, since }$\boldsymbol{J}$\emph{ is a
time like }$1$\emph{-form field there must be }(\emph{at least})\emph{ one
coordinate system where }$\boldsymbol{J}=J^{0}:=\rho_{q}\vartheta_{0}$\emph{
and thus }$A_{\mu}J^{\mu}=A_{0}\rho_{q}$\emph{. Consequently we have that }%
\begin{equation}
\int A_{\mu}J^{\mu}(-\det\boldsymbol{g})^{\frac{1}{2}}d^{4}x=Q\int A_{0}%
dx^{0}, \label{Q}%
\end{equation}
\emph{where }$Q$\emph{ is the total charge in space and then }%
$\underset{g}{\boldsymbol{\delta}}\int A_{\mu}J^{\mu}((-\det g)^{\frac{1}{2}%
})^{\frac{1}{2}}d^{4}x$\emph{ }$=0$.\medskip

Based on the last remark, we proceed with the building of our theory by
postulating a Lagrangian for the gravitational and electromagnetic fields and
their sources which must include an \textit{electromagnetic current}
$\boldsymbol{J}$ and a \textit{material medium} carrying that current (which
as in GR cannot be geometrized) as%

\begin{equation}
L:=\frac{-c^{3}}{16\pi G}(R+\frac{8\pi C}{c}A_{\mu}J^{\mu})+L_{m}. \label{22}%
\end{equation}

Thus the total action for our theory is
\begin{align}
S_{t}  &  =S+S_{m}\nonumber\\
&  =\frac{-c^{3}}{16\pi G}\int(\bar{R}+CF_{\mu\nu}F^{\mu\nu}+\frac{16\pi C}%
{c}A_{\mu}J^{\mu})(-\det\boldsymbol{g})^{\frac{1}{2}}d^{4}x\nonumber\\
&  +\frac{1}{c}\int L_{m}\left(  -\det\boldsymbol{g}\right)  ^{\frac{1}{2}%
}d^{4}x,
\end{align}
where $G$ is the gravitational constant and we require as usual in field
theories that the equations of motion of the theory are giving by%

\begin{equation}
\boldsymbol{\delta}(S+S_{m})=0\text{.} \label{29}%
\end{equation}

Now, the energy-momentum of the material charge distribution is defined by
\cite{40}%

\begin{align}
\underset{g}{\boldsymbol{\delta}}S_{m}  &  :=\frac{1}{c}\int\widetilde{T}%
_{\mu\nu}\delta g^{\mu\nu}(-\det\boldsymbol{g})^{1/2}d^{4}x,\nonumber\\
\frac{1}{2}\widetilde{T}_{\mu\nu}(-\det\boldsymbol{g})^{\frac{1}{2}}  &
=\frac{\partial\lbrack(-\det g)^{1/2}L_{m}]}{\partial g^{\mu\nu}}%
-\frac{\partial}{\partial x^{\lambda}}\frac{\partial\lbrack(-\det
g)^{1/2}L_{m}]}{\frac{\partial}{\partial x^{\lambda}}g^{\mu\nu}}, \label{30}%
\end{align}
and performing the $\underset{g}{\boldsymbol{\delta}}$ variation of
$(S+S_{m})$ we get:
\begin{equation}
\bar{G}_{\mu\nu}=\bar{R}_{\mu\nu}-\frac{1}{2}g_{\mu\nu}\bar{R}=\frac{8\pi
G}{c^{4}}(T_{\mu\nu}+\widetilde{T}_{\mu\nu})\text{ ,} \label{31}%
\end{equation}
where $\bar{G}_{\mu\nu}$ are the components of the Einstein tensor associated
with the Levi-Civita connection.

Eq.(\ref{31}) (Einstein equation in components form) gives the well know
relation between the Einstein tensor and the energy-momentum tensor of the
electromagnetic plus the energy-momentum tensor of matter on a Lorentzian
spacetime \cite{40}.

Varying Eq.(\ref{22}) with respect to and $A_{\mu}$ gives, as well known
\cite{40}, Eq. (\ref{15}), the non homogeneous Maxwell equations\footnote{The
homogeneous Maxwell equations follows trivially form $\boldsymbol{F}%
=d\boldsymbol{A}$.}, and this completes the proof of our claim that with a
special Riemann-Cartan connection it is possible to present the
electromagnetic and gravitational fields as parts of the Riemann-Cartan
structure $\langle M,\boldsymbol{g},\nabla,\tau_{\boldsymbol{g}}%
,\uparrow\rangle$ with a Lagrangian giving by Eq.(\ref{22})

\section{Lorentz Force Equation}

For simplicity, we will consider in what follows, a continuous distribution of
non-interacting incoherent charged matter, or \textquotedblleft
dust\textquotedblright\ as the material support of the electromagnetic current
$\boldsymbol{J}$. Let $\boldsymbol{\tilde{T}=}\rho_{0}c^{2}%
\boldsymbol{V\otimes V\in}\sec T_{0}^{2}M$ be the energy momentum of the
charged \textquotedblleft dust\textquotedblright\ where $\boldsymbol{V}%
=V^{\mu}\vartheta_{\mu}$ is the $1$-form velocity field of the dust
($g(\boldsymbol{V},\boldsymbol{V})=1$) and $\rho_{0}$ is its proper charged
mass density.

We calculate now the components of the covariant Riemann-Cartan covariant
derivative of $(\boldsymbol{T+\tilde{T}})$, i.e.,
\begin{equation}
\nabla_{\mu}(T^{\mu\nu}+\tilde{T}^{\mu\nu})=\bar{\nabla}_{\mu}(T^{\mu\nu
}+\tilde{T}^{\mu\nu})+K_{\mu\delta}^{\text{ \ \ }\mu}(T^{\delta\nu}+\tilde
{T}^{\delta\nu})+K_{\mu\delta}^{\ \ \nu\ }(T^{\mu\delta}+\tilde{T}^{\mu\delta
}). \label{35}%
\end{equation}

Since from Eq.(\ref{31}) it is:
\begin{equation}
\bar{\nabla}_{\mu}\bar{G}^{\mu\nu}=\frac{8\pi G}{c^{4}}\bar{\nabla}_{\mu
}(T^{\mu\nu}+\tilde{T}^{\mu\nu})=0, \label{36}%
\end{equation}

we get recalling Eq.(\ref{20}) that%
\begin{equation}
K_{\mu\delta\cdot}^{\cdot\cdot\mu}T^{\delta\nu}+K_{\mu\delta\cdot}^{\cdot
\cdot\nu\ }T^{\mu\delta}=0. \label{38}%
\end{equation}

Then, we can write
\begin{equation}
\nabla_{\mu}(T^{\mu\nu}+\tilde{T}^{\mu\nu})=K_{\mu\delta\cdot}^{\cdot\cdot\mu
}\tilde{T}^{\delta\nu}+K_{\mu\delta\cdot}^{\cdot\cdot\nu\ }\tilde{T}%
^{\mu\delta}. \label{39}%
\end{equation}

Now, recall that from Maxwell equations it follows trivially that
\begin{equation}
\nabla_{\mu}T^{\mu\nu}=\frac{1}{c}F^{\mu\nu}J_{\mu}. \label{40}%
\end{equation}

Using this result in Eq.(\ref{39}) we have
\begin{equation}
\rho_{0}c^{2}V^{\mu}\nabla_{\mu}V^{\nu}+V^{\nu}\nabla_{\mu}(\rho_{0}%
c^{2}V^{\mu})=-\frac{1}{c}F^{\mu\nu}J_{\mu}+\frac{G}{c^{4}}\rho_{0}c^{2}%
A_{\mu}V^{\delta}(F_{\delta\cdot}^{\cdot\mu}V^{\nu}+F_{\delta\cdot}^{\cdot\nu
}V^{\mu}). \label{41}%
\end{equation}

Due to $V_{\mu}V^{\mu}=1$ and the skew symmetry of $F_{\mu\nu}$, we get
contracting Eq. (\ref{41}) with $V_{\nu}$ that%
\begin{equation}
\nabla_{\mu}(\rho_{0}c^{2}V^{\mu})=\frac{G}{c^{4}}\rho_{0}c^{2}A_{\mu}%
F_{\nu\cdot}^{\cdot\mu}V^{\nu}. \label{42}%
\end{equation}

With Eq.(\ref{42}), we have from Eq.(\ref{41}),
\begin{equation}
\rho_{0}c^{2}V^{\mu}\nabla_{\mu}V^{\nu}=-\frac{1}{c}F^{\mu\nu}J_{\mu}+\frac
{G}{c^{4}}\rho_{0}c^{2}A_{\mu}V^{\mu}F_{\delta\cdot}^{\cdot\nu}V^{\delta}.
\label{43}%
\end{equation}

But for each integral line $\sigma$ (parametrized by proper time $s$) with
tangent vector field $\sigma_{\ast s}$ at $\sigma(s)$ of the flow defined by
the velocity field $\boldsymbol{V}$ we can write (with $\boldsymbol{g}%
(\sigma_{\ast s},$ $)=\left.  \boldsymbol{V}\right\vert _{\sigma}$)
\begin{equation}
V^{\mu}\nabla_{\mu}V^{\nu}=\frac{dV^{\nu}}{ds}+\bar{\Gamma}_{\mu\delta\cdot
}^{\cdot\cdot\nu}V^{\mu}V^{\delta}+K_{\mu\delta\cdot}^{\cdot\cdot\nu}V^{\mu
}V^{\delta}, \label{44}%
\end{equation}
and then, substituting Eq.(\ref{44}) in Eq.(\ref{43}) and using Eq.(\ref{20}),
we obtain
\begin{equation}
\frac{dV^{\nu}}{ds}+\bar{\Gamma}_{\mu\delta\cdot}^{\cdot\cdot\nu}V^{\mu
}V^{\delta}-\frac{\rho_{q}}{\rho_{0}c^{2}}F_{\mu\cdot}^{\cdot\nu}V^{\mu
}=0\text{ ,.} \label{45}%
\end{equation}
where have used that $J^{\mu}=c\rho_{q}V^{\mu}$ with the rest charge density
given by $\rho_{q}$.

Eq.(\ref{45}) is then identified as the Lorentz force law on a Lorentzian
spacetime \cite{40}. Note also that evaluating explicitly the covariant
derivative Eq.(\ref{42}), with the aid of Eq.(\ref{6}), we get
\begin{equation}
\bar{\nabla}_{\mu}(\rho_{0}c^{2}V^{\mu})=\partial_{\mu}[(-g)^{\frac{1}{2}}%
\rho_{0}c^{2}V^{\mu}]=0\text{ ,} \label{46}%
\end{equation}
i.e., in this model we have matter conservation.\medskip

\noindent\textbf{Remark 4\ }We observe here that if we model the matter as a
Dirac-field living in the Riemann-Cartan background we may obtain like in
\cite{popla} that the torsion tensor is also a source of the spin density.
This will be discussed elsewhere.

\section{The Gauge Invariance}

Recall that in our model the Lagrangian (excluding the electromagnetic
coupling of the electromagnetic potential with the electromagnetic current)
for the gravitational plus the electromagnetic field is geometrized, i.e., it
is given by the scalar curvature of the Riemann-Cartan connection according to
Eq.(\ref{19})
\begin{equation}
\bar{R}+\frac{G}{c^{4}}F_{\mu\nu}F^{\mu\nu}=R-\frac{8\pi G}{c^{4}}A_{\mu
}J^{\mu}.
\end{equation}

Now, we investigate what happens if we make a gauge transformation
$\boldsymbol{A\mapsto A}+d\varphi$ in the definition of the contorsion. That
transformation changes $\boldsymbol{K\mapsto}\boldsymbol{\text{\c{K}}}\ $
where the components of $\boldsymbol{\text{\c{K}}}$ are
\begin{equation}
\text{\c{K}}_{\mu\nu\cdot}^{\cdot\cdot\ \lambda}=K_{\mu\nu\cdot}^{\cdot
\cdot\lambda}+\frac{G}{c^{4}}(\partial_{\mu}\varphi)F_{\nu\cdot}^{\cdot
\lambda}\text{ }. \label{47}%
\end{equation}

Then, we get an new Riemann-Cartan curvature tensor whose components are%
\begin{equation}
\text{\c{R}}_{\mu\nu\lambda\cdot}^{\cdot\cdot\cdot\chi}=R_{\mu\nu\lambda\cdot
}^{\cdot\cdot\cdot\chi}+\frac{G}{c^{4}}[\bar{\nabla}_{\mu}(F_{\lambda\cdot
}^{\cdot\chi}\partial_{\nu}\varphi)-\bar{\nabla}_{\nu}(F_{\lambda\cdot
.}^{\cdot\chi}\partial_{\mu}\varphi)]\label{48}%
\end{equation}

From Eq.(\ref{48}), the new scalar curvature \c{R} is
\begin{equation}
\text{\c{R}}=R+\frac{8\pi G}{(-g)^{\frac{1}{2}}c^{5}}\partial_{\mu
}[(-g)^{\frac{1}{2}}\varphi J^{\mu}]\text{.} \label{49}%
\end{equation}

Then since $($\c{R}$-R)$ differs by an exact differential, if we take \c{R} as
the new Lagrangian for the electromagnetic plus gravitational fields we get
the same equations of motion as before.

We conclude that the freedom in choosing an electromagnetic gauge in our
physical equations means the freedom, within a gauge, to choose
the\ Riemann-Cartan curvature tensor \c{R}$_{\mu\nu\lambda\cdot}^{\cdot
\cdot\cdot\chi}$ (through \c{K}$_{\mu\nu\cdot}^{\cdot\cdot\lambda}$). So,
there is a class of gauge equivalent Riemann-Cartan structure describing the
same gravitational plus electromagnetic field.

\section{Conclusions.}

We have obtained a set of equations, namely Einstein equations, Maxwell
equations and the Lorentz force equations describing a continuous distribution
of charged matter interacting with the electromagnetic and gravitational
fields from a \textit{geometric} point of view. We note, however, that those
equations have the same form as if they were written on a Lorentzian
spacetime, although we have postulated as Lagrangian for the free
electromagnetic plus gravitation field the scalar curvature of a (particular)
Riemann-Cartan spacetime. Then, if there is an electromagnetic field generated
by a current distribution on a Lorentzian spacetime modeling a gravitational
field we can think of these two fields as geometrical properties of a
particular Riemann-Cartan spacetime structure, the one whose contortion tensor
is given by Eq.(\ref{20}), although this is not apparent in the usual physical
equations. Also the contortion tensor of our theory encodes the same
information as the one that is contained in the Chern-Simons term $A\wedge
dA$, which is proportional to the spin density of the electromagnetic field.
Finally we observe that despite the fact that the contortion tensor and the
Riemann-Cartan curvature tensor of our theory is not gauge invariant, the
resulting field equations obtained through the complete Lagrangian involving
the coupled interacting system consisted of gravitational field plus the
electromagnetic fields and the charge current produce equations for those
fields and equations of motion of the charged matter\ that are gauge
invariant, as they should be. So, for each gauge we have a different, but
\textit{equivalent} Riemann-Cartan spacetime structure defined by another gauge

\bigskip\noindent\textbf{Acknowledgments\ }\newline Authors dedicate this
paper to the memory of Professor Jaime Keller. They also would like to thank
J. Vaz, Jr., M. A. Faria Rosa and R. da Rocha for useful discussions.


\begin{thebibliography}{99}                                                                                               %


\bibitem {22}Adler, R., Bazin, M., and Schiffer, M. \textit{Introduction to
General Relativity}, chapter 13, McGraw-Hill, New York, 1965.\ 

\bibitem {32}Andrade,V.C., and Pereira, J. G., Torsion and the Electromagnetic
Field., \textit{Int. J. Mod. Phys.} \textit{D }\textbf{8}, 141-151(1999)

\bibitem {7}Appelquist, T., Chodos, A., and Freund, P.G.O., \textit{Modern
Kaluza-Klein Theories, }Addison-Wesley, New York, 1987.

\bibitem {babak}Babak, S. V., and Grishckuk, L. P., The Energy Momentum Tensor
for the Gravitational Field., \textit{Phys. Rev. D }\textbf{61}, 024038.1-18
(1999) [arXiv: gr-qc/9907027 v2]

\bibitem {43}Barut, A. O., \textit{Elecrodynamics and Classical Theory of
Fields and Particles, }Dover, New York, 1980.

\bibitem {19}Beil, R. G. Electrodynamics from a Metric\textit{, Int. J. Theor.
Phys.} \textbf{26}, 189-197 (1986).

\bibitem {11}Borchsenius, K., An extension of the Nonsymmetric Unified Field
Theory, \textit{Gen. Rel. Grav}. \textbf{7}, 527- 534 (1976).

\bibitem {39}Capozzielo, S., Lambiase, G., Stornailo, C. Geometric
Classification of Torsion Tensor of Space-Time, \textit{Ann. Phys}.
\textbf{10},713-727(2001) \texttt{[arXiv: gr-qc 0101038 v1]}

\bibitem {23}Cartan, E., \textit{On Manifolds with an Affine Connection and
the Theory of General Relativity}, Bibliopolis, Napoli, 1986.

\bibitem {chyba}Chyba, C.F., Kaluza-Klein Unified Field Theory And Apparent
Four-Dimensional Space-Time, \textit{Am. J. Phys.} \textbf{53}, 863-872 (1985)

\bibitem {choquet}Choquet-Bruhat, Y., DeWiit-Morette, C, \ and
Dillard-Bleick,\ M., \textit{Analysis, Manifolds and Physics} (revised
edition), North-Holland, Amsterdam, 1982.

\bibitem {8}Coquereaux, R., and Jadczyk, A. \textit{Riemannian Geometry, Fiber
Bundles, Kaluza-Klein Theories And} \textit{All That}, Lectures Notes in
Physics\textbf{ 16}, World Scientific, Singapore, 1988.

\bibitem {cg}Cottingham, W. N., and Greenwood, D. A., \textit{An Introduction
to the Standard Model of Particle Physics} (second edition), Cambridge Univ.
Press, Cambridge, 2007.

\bibitem {rr2010}da Rocha, R. and Rodrigues, W. A. Jr. Pair and Impar, Even
and Odd Differential Forms and Electromagnetism, \textit{Ann. Phys. }(Berlin)
\textbf{19}, 6-34 (2010).

\bibitem {4}Eddington, A. S., \textit{The Mathematical Theory of Relativity,
}3th. ed., Chelsea, New York, 1975.

\bibitem {17}Einstein, A., \textit{The Meaning of Relativity, }appendix
II\textit{, }Princeton University Press, Princeton, 1956.

\bibitem {ferod2010}Fern\'{a}ndez, V. V. and Rodrigues, W. A. Jr.,
\textit{Gravitation as a Plastic Distortion of the Lorentz Vacuum},
Fundamental Theories of Physics \textbf{168, }Springer, Heidleberg,\textbf{ }2010.

\bibitem {18}Fontaine, M., and Amiot, P., Geometrical Aspects of
Electromagnetism, I., \textit{Ann. Phys.} \textbf{147}, 269-280 (1983).

\bibitem {30}Fradkin, E. S., and Tseytlin, A. A., Effective Field Theory from
Quantized Strings, \textit{Phys. Lett. B}\textbf{158}, 316-322 (1985).

\bibitem {37}de Andrade, L. C. G., and \ Lopes, M, Detecting Torsion from
Massive Electrodynamics, \textit{G. Rel. Grav.} \textbf{25,} 1101-1106. (1993).

\bibitem {devries}de Vries, H., On the electromagnetic Chern Simons spin
density as hidden variable and EPR correlations.
\texttt{[http://physics-quest.org/ChernSimonsSpinDensity.pdf]}

\bibitem {34}Garecki, J., Is Torsion Needed in Theory of Gravity ?,\textit{
Rel. Grav. Cosmol. }\textbf{1}, 43-59 (2004)\textit{ }

\texttt{[arXiv: gr-qc/0103029 v1]}

\bibitem {gs}G\"{o}ckeler, M, and Sch\"{u}cker, T., Difeerentil Geoemtry,
Gauge, Gauge Theories, and Gravity, Cambridge University Press, Cambridge, 1987.

\bibitem {15}Hammond, R. T., Gravitation, Torsion, and Electromagnetism,
\textit{Gen. Rel. Grav.} \textbf{20}, 813-827 (1988).

\bibitem {25}Hammond, R. T., Spin, Torsion, Forces, \textit{Gen. Rel. Grav}.
\textbf{26}, 247-263 (1994).

\bibitem {31}Hammond, R. T., New Fields in General Relativity, \textit{Cont.
Phys}. \textbf{36}, 103-114 (1995).

\bibitem {33}Hammond, R. T., Upper Limit on the Torsion Coupling Constant,
\textit{Phys. Rev. D}\textbf{52}, 6918-6921 (1995).

\bibitem {24}Hehl, F. W., von der Heyde, P., and Kerlick, G., General
Relativity with Spin and Torsion: Foundations and Prospects, \textit{Rev. Mod.
Phys}. \textbf{48}, 393-416 (1976).

\bibitem {25aa}Hehl, F. W., Spin and Torsion in General Relativity: I.
Foundations, \textit{Gen. Rel. Grav.}, \textbf{4}, 333-349 (1973).

\bibitem {25a}Hehl, F. W., and Obukhov, Y. N. How Does the Electromagnetic
Field Couple to Gravity, in Particular to Metric, Nonmetricity, Torsion and
Curvature?, \textit{Lecture Notes in Phys.} \textbf{562}, 479-504
(2001)\texttt{[arXiv: gr-qe/0001010 v2]}

\bibitem {helobu}Hehl, F. W., and Obukhov, Y. N., \textit{Foundations of
\ Classical Electrodynamics, Charge, Flux and Metric}, Birkh\"{a}user, Boston, (2003).

\bibitem {14}Jacubiec, A., and Kijowski, J., On Interaction of The Unified
Maxwell-Einstein Field with Spinorial Matter, \textit{Lett. Math. Phys}.
\textbf{9}, 1-11 (1985).

\bibitem {5}Kaluza, Th., Zum Unit\"{a}tsproblem in der
Physik,\textit{\ Sitzungsber Preuss. Akad. Wiss. }Berlin, 966-972 (1921).

\bibitem {27}Kibble, T., Lorentz Invariance and the Gravitational Field,
\textit{J. Math. Phys}. \textbf{2}, 212-221 (1961).

\bibitem {kiehn}Kiehn, R. M., \textit{Non-Equilibrium and Irreversible
Thermodynamics from a Topological Perspective}. \textit{Adventures in Applied
Topology} vol.\textbf{1}, 2008. \texttt{[http://www.lulu.com/spotlight/kiehn]}

\bibitem {6}Klein\textit{, }O., Quantentheorie und F\"{u}nfdimensionale
Relativit\"{a}tstheorie, \textit{Z. Phys}. \textbf{37}, 895-906
(1926)\textit{.}

\bibitem {38}Kleinert, H., \textit{Gauge Fields in Condensed Matter,} World
Scientific, Singapore, 1989.

\bibitem {40}Landau, L. D., and Lifshitz, E. M., \textit{The Classical Theory
of Fields, }Pergamon, Oxford, 1971.

\bibitem {29}Maluf, J. W., Hamiltonian Formulation of the Teleparallel
Description of General Relativity, \textit{J. Math. Phys}. \textbf{35},
335-343 (1994).

\bibitem {12}Moffat, J. W., Space-time Structure in a Generalization of
Gravitation Theory, \textit{Phys. Rev}. \textit{D }\textbf{15}, 3520-3529 (1977).

\bibitem {moffat}Moffat, J. W., \textit{Reinventing Gravity}, Smithsonian
Books, Harper Coolins Publ, New York, 2008.

\bibitem {13}McKellar, R. J., Asymmetric Connection Treatment of the
Einstein-Maxwell Field Equations, \textit{Phys. Rev. D }\textbf{20}, 356-361 (1979).

\bibitem {36}Nakahara, M, \textit{Geometry, Topology and Physics}, IOP Publ.
Ltd, Bristol and Philadephia, 1990

\bibitem {nrr2010}Note-Cuello, E. A., da Rocha, R. and Rodrigues, W. A. Jr.,
Some Thoughts on Geometries and on the Nature of the Gravitational Field ,
\textit{J. Phys. Math.} \textbf{2}, 20-40 (2010)

\bibitem {42}Pauli, W., \textit{Theory of Relativity}, Dover Publ. Inc., New
York, 1981.

\bibitem {popla}Poplawski N. J., Torsion as Electromagnetism and Spin,
\textit{Int. J. Theor. Phys}. \textbf{49}, 1481-1488 (2010).

\bibitem {16}Ringermacher, H., An Electrodynamic Connection, \textit{Class.
Quant. Grav.} \textbf{11}, 2383-2394 (1997).

\bibitem {rodcap2007}Rodrigues, W. A. Jr. and Capelas de Oliveira, E.,
\textit{The Many Faces of Maxwell, Dirac and Einstein Equations}, Lecture
Notes in Physics \textbf{722}, Springer, Heidleberg, 2007.

\bibitem {rodflb}Rodrigues, W. A. Jr., Differential Forms on Riemannian
(Lorentzian) and Riemann-Cartan Structures and Some Applications to Physics,
\textit{Ann. Fond. L. de Broglie,} \textbf{32 }(special issue dedicated to
torsion), 425-478 (2007).\texttt{[arXiv:0712.3067v6 [math-ph]]}

\bibitem {rod2011}Rodrigues, W. A. Jr., \textit{On the Nature of the
Gravitational Field and its Legitimate Energy-Momentum Tensor.}
\texttt{[arXiv:1109.5272[math-ph]]}

\bibitem {1}Weyl, H., Gravitation und Elektrizit\"{a}t, \textit{Sitzungsber.
Preuss. Akad. Wiss }\textbf{26}, 465-478 (1918)

\bibitem {2}Weyl, H., \textit{Space,Time, Matter}, 4th. ed.,\textit{\ }Dover
Publications, New York, 1950.

\bibitem {sauer}Sauer, T., Field Equations in Teleparallel Spacetime:
Einstein's Fernparallelismus Approach Towards Unified Field Theory.\textit{,
Historia Math. }\textbf{33}\textit{, }399-439 (2006).
\texttt{[arXiv:physics/0405142v1]}

\bibitem {sawu}Sachs, R. K. and Wu, H., \textit{General Relativity for
Mathematicians}, Springer-Verlag, New York, 1997.

\bibitem {28}Sciama, D., The Physical Structure of General Relativity,
\textit{Rev. Mod. Phys} \textbf{36}, 463--469 (1964).

\bibitem {9}Schr\"{o}dinger, E., The General Unitary Theory of the Physical
Fields, \textit{Proc. R. Irish Acad. A }\textbf{49, }43-58 (1943).

\bibitem {35}Sivaram, C., and de Andrade, L. C. G., \textit{Torsion Gravity
Effects on Carged-Particle and Neutron Interferometers.} \texttt{[arXiv:
gr-qc/0111009 v1]}

\bibitem {3}Souza, Q. A. G., and Rodrigues, W. A. Jr, The Dirac Operator and
the Structure of Riemann-Cartan-Weyl Spaces,. in Letelier, P., and Rodrigues,
W. A. Jr. (eds.), \textit{Gravitation: The Spacetime Structure, }pp. 179-212,
World Scientific, Singapore, (1994).

\bibitem {26}Utiyama, D., Invariant Theoretical Interpretation of Interaction,
\textit{Phys. Rev. }\textbf{101}, 1597-1607 (1956).

\bibitem {20}Vargas, J. G., Geometrization of the Physics with Teleparallelism
I. The Classical Interactions., \textit{Found. Phys}. \textbf{22}, 507-526 (1992)

\bibitem {21}Vargas, J. G., Torr, D. G., and Lecompte, A.,Geometrization of
the Physics with Teleparallelism. II. Towards a Fully Geometric Dirac
Equation, \textit{Found. Phys.} \textbf{22}, 527-547 (1992).
\end{thebibliography}
\end{document}